\documentstyle[floats,prd,aps,epsf,eqsecnum,12pt]{revtex}

\makeatletter
\newbox\tempboxa
\newdimen\captionboxsubcount
\def\capsize#1{\captionboxsubcount=#1pt}
\newdimen\captionboxsub
\captionboxsub=\hsize \advance\captionboxsub by -\captionboxsubcount
\advance\captionboxsub by -\captionboxsubcount
\long
\def\@makecaption#1#2{
\setbox\@tempboxa\hbox{#1 #2}
\ifdim \wd\@tempboxa >\captionboxsub
\rightskip=\captionboxsubcount \leftskip=\captionboxsubcount #1 #2
\else \hbox to\hsize{\hfil\box\@tempboxa\hfil}
\fi}
\makeatother
\capsize{30}

\begin{document}

\begin{titlepage}
\begin{flushright}
\begin{minipage}{5cm}
\begin{flushleft}
\small
\baselineskip = 13pt
LA-UR-99-4856\\ 
TMUP-HEL-9911 \\
YCTP-P23-99\\ 
hep-ph/9909296 \\
\end{flushleft}
\end{minipage}
\end{flushright}
\begin{center}
\Large\bf
A Two-dimensional Model with Chiral Condensates and Cooper Pairs having QCD-like Phase Structure
\end{center}
\vfil

\begin{center}

Alan {\sc Chodos}\footnote{ Electronic address : {\tt
chodos@hepvms.physics.yale.edu}}\\ { \it \qquad Department of
Physics, Yale University, New Haven, CT 06520-8120}\\

Fred {\sc Cooper}\footnote{Electronic address: {\tt
cooper@schwinger.lanl.gov}}
\\ {\it \qquad Los Alamos National Laboratory, Los Alamos, NM 87545} \\

Wenjin {\sc Mao}\footnote{Electronic address: {\tt
maow@physics.bc.edu}} \\
{\it \qquad Department of Physics, Boston College, Chestnut Hill, MA
02467} \\

Hisakazu {\sc Minakata}\footnote{Electronic address: {\tt
minakata@phys.metro-u.ac.jp}} \\
{\it \qquad Department of Physics, Tokyo Metropolitan University\\
Minami-Osawa, Hachioji, Tokyo 192-0397, Japan and\\
Research Center for Cosmic Neutrinos, Institute for Cosmic Ray
Research\\
University of Tokyo, Tanashi, Tokyo 188-8502, Japan}

Anupam {\sc Singh}\footnote{Electronic address: {\tt
singh@lanl.gov}} \\
{\it \qquad Los Alamos National Laboratory, Los Alamos, NM 87545} \\

\vskip .5cm
 {\sc }

\qquad

\end{center}
\vfill
\newpage
\begin{center}
\bf
Abstract
\end{center}
\begin{abstract} 

We generalize our previous model[9] to  an O(N) symmetric 
two-dimensional model which possesses chiral symmetry breaking 
($\langle \bar{\psi}\psi \rangle$ condensate) and superconducting 
(Cooper pair $\langle \psi \psi \rangle$ condensates) phases at large-N. 
At zero temperature and density, the model can be solved analytically 
in the large-N limit. We perform the renormalization explicitly and obtain 
a closed form expression of the effective potential. 
There exists a renormalization group invariant parameter $\delta$
that determines which of the $\langle \bar{\psi}\psi \rangle$ ($\delta > 0$) 
or $\langle \psi \psi \rangle$ ($\delta < 0$) condensates exist in the 
vacuum.
At finite temperatures and densities, we map out the phase structure
of the model by a detailed numerical analysis of the renormalized 
effective potential. For $\delta$ positive and sufficiently large, the 
phase diagram in the $\mu$-$T$ 
(chemical potential-temperature) plane exactly mimics the features 
expected for QCD with two light flavors of quarks.
At low temperatures there exists low-$\mu$ chiral symmetry breaking 
and high-$\mu$ Cooper pair condensate regions which are separated by a
first-order phase transition. At high $\mu$, when the temperature is raised, 
the system undergoes a second-order phase transition from the superconducting 
phase to an unbroken phase in which both condensates vanish. For a range of 
values of $\delta$ the theory 
possesses a tricritical point ($\mu_{tc}$ and $T_{tc}$); for 
$\mu > \mu_{tc}$ ($\mu < \mu_{tc}$) the phase transition from the 
low temperature chiral symmetry breaking phase to unbroken phase is 
first-order (second-order).
For the range of $\delta$ in which the system mimics QCD, we expect
the model to be useful for the investigation of dynamical aspects of
nonequilibrium phase transitions, and to provide information relevant
to the study of relativistic heavy ion collisions and the dense interiors of
neutron stars.

\baselineskip = 17pt

\end{abstract}
\begin{flushleft}

PACS numbers:
\end{flushleft}
\vfill
\end{titlepage}

\section{Introduction}

\bigskip

The phase structure of $QCD$ at non-zero  temperature and baryon density
is important for the physics of neutron stars and heavy ion collisions.

An approximate phase structure for QCD with two massless quarks
has been mapped out in various mean field and perturbative approximations 
and a rich
structure has emerged.  For a recent review, see
\cite{ref:review}. In addition to the well known chiral symmetry broken 
and restored phases recent investigations have revealed the possibility 
of a color
superconducting phase at low temperatures and relatively high densities
\cite{ref:super1} \cite{ref:super2} \cite{ref:super3}.
In the chiral condensation regime at zero chemical potential, the
phase transition to the unbroken mode is second order as we raise the
temperature \cite{ref:PW84}.
As we increase the chemical potential at fixed low temperature  there is
a first order transition to a superconducting phase. There is also a
regime where as we increase the temperature the phase transition from
the chirally broken phase to the unbroken phase is first order, so that
somewhere along the line separating these phases there is a tricritical
point \cite{ref:tricritical}.
These results are summarized in \cite{ref:review}.

  One of the more interesting questions is what happens in a
dynamical situation such as a heavy ion collision, in which the system
traverses the
various
phase transitions as it expands and cools. One would like to study
the correlation
functions  to see whether there is qualitatively different behavior in
crossing  the first order or second order transition regions as a function of
the proper evolution time and whether this difference would lead
to some interesting experimental signatures at RHIC.

In order to get a better
handle on this latter question, we  propose a simple 
1+1 dimensional
model which contains several of the features of two flavor massless QCD
in mean
field approximation
(similar phase structure and asymptotic freedom) which would lend itself
to
dynamical computational simulations pertinent to the (one dimensional) 
expansion of a Lorentz contracted disc of quark matter.  Thus we hope to
explore the behavior of a system of ``quarks" evolving through first
and second order phase transitions to either a final state with chiral
condensates or to a superconducting final state.  This calculation would
be
similar in spirit to those that were done in exploring the chiral phase
transition in
the linear sigma model \cite{ref:DCC}.

 In this  paper we restrict
ourselves to
mapping out the phase structure of our model at finite temperature and
chemical
potential in  a large-$N$  approximation, for
use in obtaining initial conditions for our future dynamical
calculations. 
Our simple model  combines the Gross-Neveu model
\cite{ref:GN}  with a
model for Cooper pairs that we introduced recently \cite{ref:paper1}. It
turns out to have
many of the features of QCD that we ultimately want to capture in more
realistic calculations.  Namely, the theory in the Gross-Neveu sector
has the
same second-order, first-order,  tricritical point behavior for the chiral
condensate $< \bar{q} q>$ as a function of
temperature \cite{ref:GN2} and chemical potential \cite{ref:GN2} \cite{ref:minakata} as
QCD with two massless flavors.
Adding the second
interaction
also adds a new phase where there is superconductivity at some finite
chemical
potential as also expected in QCD with two massless flavors. The model
has  a
well defined $1/N$ expansion and is asymptotically free so that it does
not
suffer from the cutoff dependences of $3+1$ dimensional effective field
theories considered by others \cite{ref:super1} \cite{ref:super2}.

Thus we will be investigating a 1+1 dimensional model
governed by two independent couplings: ~the original Gross-Neveu term
\cite{ref:GN} that promotes the condensation of $\langle \bar{q} q
\rangle$, and the term considered in our earlier paper
\cite{ref:paper1}that
produces a $\langle qq \rangle$ condensate. We first determine the
unrenormalized effective potential at leading order in a large-N
expansion,
by introducing collective coordinates for the $\bar{q} q
$, and  $qq$ operators,
and integrating out the fermions in the usual fashion by using the
Hubbard-Stratonovich trick \cite{ref:hub}.

At zero temperature and chemical potential  we  derive
a closed-form analytic expression for the renormalized effective
potential.
 We 
find that there is one dimensionless parameter $\delta$, independent of
the
renormalization scale, whose value determines which of the
condensates is present. This situation might be described as
``partial dimensional transmutation": the unrenormalized theory has
two bare couplings whereas the renormalized one has a
renormalization scale, which is arbitrary (but which can be related to 
the physical fermion or Cooper pair gap mass) , and a
dimensionless parameter $\delta$ independent of this scale, that controls the
physics.
We find that the gap equations have three types of solution: ~two in
which one or the other of the condensates vanish, and a third,
mixed case, in which both condensates are non-vanishing. It turns
out, however, that the true minimum of the effective potential is
at a point where one of the condensates vanishes except at the
particular
point $\delta=0$ (to be discussed further below)where one is at
a first order phase transition so that phase coexistence can occur.
Depending
on the sign of $\delta$ we therefore have 
very different behavior at zero temperature and chemical
potential-- namely either chiral condensates or Cooper-pairs being
present.
 
The phase structure of the limiting theories which have only one
coupling
constant is easy to determine analytically, and we include for
completeness
a discussion of these two particular cases which benchmark our numerical
study of the general case.   Performing the integration in our 
expression for the renormalized effective potential numerically, we
then map out the phase diagram for the more general two coupling
constant case
as a  function of $\delta$.  We find a regime  of  (positive) $\delta$
which
remarkably mimics the phase structure of two flavor QCD described above.
It has
a tricritical point as well as a first order phase transition as a 
function of chemical potential from the chirally broken phase into the
superconducting phase.  We determine  the tricritical point and the
critical temperature at which  the superconducting phase transition
occurs
as a function of $\delta$.  As we decrease  the magnitude of $\delta$,
first
the regime of first order phase transition from the chiral phase
disappears,
and then at the special point $\delta=0$ the possibility of a chiral
symmetry
broken phase totally disappears so that when $\delta < 0$ one only has
the possibility for a  superconducting broken symmetry mode at low
temperatures.

\section{General Considerations}

\bigskip
We consider the most general Lagrangian with quartic fermion couplings, possessing $O(N)$ flavor symmetry and discrete chiral symmetry.

\begin{eqnarray}
{\cal L} &=& \bar{\psi}^{(i)} i \bigtriangledown  \!\!\!\!\!\! /
\psi^{(i)} + {1
\over 2} g^2 [\bar{\psi}^{(i)} \psi^{(i)}][\bar{\psi}^{(j)} \psi^{(j)}]
\nonumber \\
&+& 2 G^2 (\bar{\psi}^{(i)} \gamma_5 \psi^{(j)})(\bar{\psi}^{(i)}
\gamma_5 \psi^{(j)}) - \mu \psi^{\dagger (i)} \psi^{(i)} .
\end{eqnarray}

\noindent
The flavor indices, summed on from $1$ to $N$, have been explicitly
indicated. The first quartic term is the usual Gross-Neveu
interaction, whereas the second such term, which differs in the
arrangement of its flavor indices, induces the pairing force to
leading order in ${1 \over N}$. This term is possible because we
demand only $O(N)$ symmetry as opposed to the $SU(N)$ symmetry of the original Gross-Neveu model. In the final term, $\mu$ is the
chemical potential.

Strictly speaking, a $\langle \psi \psi \rangle$ condensate cannot
form, because it breaks the $U(1)$ of Fermion number and hence
violates Coleman's theorem  \cite{ref:mermin}. Similarly,
$\langle \bar{\psi} \psi \rangle$ as well as $\langle \psi \psi \rangle$
condensates cannot exist at finite temperature in one spatial dimension
because of the Mermin-Wagner theorem  \cite{ref:mermin}.  Nevertheless, it is
meaningful to study the formation of such condensates to leading
order in ${1 \over N}$, as explained in ref. \cite{ref:Witten}.
%

Our conventions are: ~$\gamma^0 = \sigma_1$; $\gamma^1 = - i
\sigma_2$; $\gamma_5 = \sigma_3$. The pairing term, proportional to
$G^2$,
may then be rewritten:

\begin{eqnarray}
2G^2 \bar{\psi}^{(i)} \gamma_5 \psi^{(j)} \bar{\psi}^{(i)} \gamma_5
\psi^{(j)} = - G^2 [\epsilon_{\alpha\beta} \psi_{\alpha}^{\dagger (i)}
\psi_{\beta}^{\dagger (i)}]
[\epsilon_{\gamma \delta} \psi_{\gamma}^{(j)}
\psi_{\delta}^{(j)}] ~.
\end{eqnarray}

\bigskip\noindent
Following standard techniques [8] we add the following terms
involving auxiliary fields $m$, $B^{\dagger}$, and $B$:

\begin{eqnarray}
\bigtriangleup {\cal L} = - {1 \over 2 g^2} [m + g^2 \bar{\psi} \psi]^2
-
{1 \over G^2} (B^{\dagger} - G^2 \epsilon_{\alpha\beta}
\psi_{\alpha}^{\dagger (i)} \psi_{\beta}^{\dagger (i)})(B + G^2
\epsilon_{\gamma \delta} \psi_{\gamma}^{(j)} \psi_{\delta}^{(j)}) ~.
\end{eqnarray}

\noindent
This addition to ${\cal L}$ will not affect the dynamics. In ${\cal
L}^{\prime}
= {\cal L} + \bigtriangleup {\cal L}$, the terms quartic in fermion
fields cancel, and we have

\begin{eqnarray}
{\cal L}^{\prime} = \bar{\psi} (i \bigtriangledown  \!\!\!\!\!\! /
- m - \mu
\gamma^0) \psi - {m^2 \over 2 g^2} - {B^{\dagger}B \over G^2} + B
\epsilon_{\alpha\beta} \psi_{\alpha}^{\dagger (i)}
\psi_{\beta}^{\dagger (i)} - B^{\dagger} \epsilon_{\alpha\beta}
\psi_{\alpha}^{(i)} \psi_{\beta}^{(i)} ~.
\end{eqnarray}

\bigskip\noindent
We integrate out $\psi$ and $\psi^{\dagger}$ to obtain the
effective action depending on the auxiliary fields $m$, $B$ and
$B^{\dagger}$:

\begin{eqnarray}
\Gamma_{eff} (m, B, B^{\dagger}) = \int d^4x (- {m^2 \over 2 g^2} -
{B^{\dagger}B \over G^2}) - {i \over 2} Tr \ln A^T A - {i \over 2} Tr
\ln
[{\bf 1} + M^2 (A^T)^{-1} \sigma_2
A^{-1} \sigma_2]  \label{eq:Gamma}
\end{eqnarray}

\bigskip\noindent
where we have subtracted a constant (independent of the auxiliary
fields) and have defined:

\begin{eqnarray}
A = \gamma^0 (i \bigtriangledown  \!\!\!\!\!\! / - m - \mu
\gamma^0) = i \partial_0 + i \sigma_3 \partial_x - \mu - m \sigma_1
\end{eqnarray}

\bigskip\noindent
so that $A^T = - i \partial_0 - i \sigma_3 \partial_x - \mu - m
\sigma_1$.

Since we are looking for a vacuum solution, we have assumed in
(\ref{eq:Gamma}) that $B, B^{\dagger}$ and $m$ are constants and have set $M^2
= 4 B^{\dagger}B$. The trace on flavor indices will give a factor
$N$. The large-$N$ limit is achieved by setting $g^2N
= \lambda$, and $G^2N = \kappa/4$, and letting $N \rightarrow \infty$
with
$\lambda$ and $\kappa$ fixed.
We define the effective potential $V_{eff}$ via

\begin{eqnarray}
\Gamma_{eff} = - N(\int d^2x) V_{eff}
\end{eqnarray}

\bigskip\noindent
and we therefore have

\begin{eqnarray}
V_{eff} (m, M) = {m^2 \over 2 \lambda} + {M^2 \over \kappa} +
V_{eff}^{(1)} (m, M) ~,
\end{eqnarray}

\bigskip\noindent
with $V_{eff}^{(1)} (m, M) = {i \over 2} [tr \ln (A^TA)_{xx} + tr
\ln ({\bf 1} + M^2(A^T)^{-1} \sigma_2 A^{-1} \sigma_2)_{xx}]$, where
now the trace is only over the spinor indices.

We next generate the local extrema of $V_{eff}$ by solving

\begin{eqnarray}
{\partial V_{eff} \over \partial m^2} = {\partial V_{eff} \over
\partial M^2} = 0 ~. \label{eq:ext}
\end{eqnarray}

\bigskip\noindent
We evaluate the matrix products in $V_{eff}^{(1)}$ in momentum
space, with $\partial_{\mu} \rightarrow i k_{\mu}$. The traces can
be done with the help of

\begin{eqnarray}
tr [{1 \over V_0 + \vec{V} \cdot \vec{\sigma}}] = {2 V_0 \over
V_0^2 - \vec{V}^2}
\end{eqnarray}

\bigskip\noindent
for any $V_0, \vec{V}$. After some manipulation, equations (\ref{eq:ext})
become

\begin{eqnarray}
{1 \over 2 \lambda} = - {\partial V_{eff}^{(1)} \over \partial m^2}
= i \int {d^2 k \over (2 \pi)^2} {[k_0^2 - k_1^2 + \mu^2 + M^2 - m^2]
\over D}  \label{eq:gngap1}
\end{eqnarray}

\begin{eqnarray}
{1 \over \kappa} = - {\partial V_{eff}^{(1)} \over \partial M^2}
= i \int {d^2 k \over (2 \pi)^2} {[k_0^2 - k_1^2 - \mu^2 - M^2 + m^2]
\over D} \label{eq:cgap1}
\end{eqnarray}

\bigskip\noindent
where 
\begin{equation}
D = [k_0^2 - k_1^2 - M^2 + m^2 - \mu^2]^2 - 4[m^2 k_0^2 + \mu^2
k_1^2 - m^2 k_1^2]  \label{eq:D}
\end{equation}
 In this expression, $k_0$ is shorthand for
$k_0 + i\epsilon sgn k_0$, where $\epsilon \rightarrow 0^+$. This
prescription correctly implements the role of $\mu$ as the chemical
potential.

The equations can be reduced further by doing the $k_0$ integral.
Let us define $k_{\pm} = \sqrt{b_1 \pm 2 b_2}$, where $b_1 = M^2  +
m^2 + \mu^2 + k_1^2$, and $b_2 = [M^2 m^2 + \mu^2(k_1^2 + m^2)]^{{1
\over 2}}$. Then evaluating the $k_0$ integral by contour methods,
taking proper account of the $i\epsilon$ prescription mentioned
above, we find

\begin{eqnarray}
{1 \over 2\lambda} = {1 \over 8\pi} \int_{- \Lambda}^{\Lambda} dk_1
[{1 \over k_+} + {1 \over k_-} + {(M^2 + \mu^2) \over \sqrt{M^2  m^2 +
\mu^2 (k_1^2 + m^2)}} ({1 \over k_+} - {1 \over k_-})]
\end{eqnarray}

\noindent
and

\begin{eqnarray}
{1 \over \kappa} = {1 \over 8\pi} \int_{- \Lambda}^{\Lambda} dk_1
[{1
\over k_+} + {1 \over k_-} + {m^2 \over \sqrt{M^2 m^2 +
\mu^2 (k_1^2 + m^2)}} ({1 \over k_+} - {1 \over k_-})] ~.
\end{eqnarray}

\bigskip\noindent
The $k_1$ integrals are logarithmically divergent and we have
regularized them by imposing a cutoff $\Lambda$. This will be
absorbed in the renormalization process to be described in the next
section. Note, however, that the combination $ \delta = {1 \over
\kappa}-{1 \over 2\lambda}
$ is given by a convergent integral. This fact will
ultimately lead to the renormalization-scale independent constant
mentioned in the introduction.

We observe from the form of equations (\ref{eq:gngap1}) and (\ref{eq:cgap1}) that the
function $V_{eff}^{(1)}$ can be reconstructed by integrating with
respect to $m^2$ and $M^2$ in the expressions for ${1 \over 2
\lambda}$ and ${1
\over \kappa}$. This will determine $V_{eff}^{(1)}$ up to a single
constant
$V_{eff}^{(1)}(0,0)$, which can be chosen arbitrarily without
affecting any physical quantity. Explicitly performing this
integration we obtain for the unrenormalized
 determinant correction to the
effective potential

\begin{equation}
V^{(1)} (m,M) = -{1 \over 2 \pi} \int_0^{\Lambda} dk_1 [ k_+ + k_{-} ]  \label{eq:v1mM}
\end{equation}

\bigskip

To generalize this discussion to the case of non-zero temperature,
one returns to eqns. (\ref{eq:gngap1}) and (\ref{eq:cgap1}), and one continues to
Euclidean space via the replacement $k_0
\rightarrow  -i k_4$ with $k_4$ now considered real.
The statistical-mechanical partition function is obtained from the
Euclidean zero temperature path integral by integrating over a
finite regime in imaginary time $\tau = it$ from $0$ to $\beta = {1
\over kT}$. Because of the cyclic property of the trace, the
Fermion Green's functions are anti-periodic in $\tau$ and one has
the replacement

\begin{equation}
\int dk_4 \rightarrow  {2 \pi \over \beta} \sum_{n} \label{eq:euc}
\end{equation}

\noindent
where the antiperiodicity gives the Matsubara frequencies:

\begin{equation}
\omega_n = k_{4_n} = {(2 n+1) \pi \over \beta}
\end{equation}

To do the sum over the Matsubara frequencies, one uses the calculus of
residues to obtain the identity:

\begin{equation}
 {2 \over \beta} \sum_n f(i \omega_n) = - \sum_s \tanh {\beta z_s \over
2}~~
 Res f(z_s)
\end{equation}

\noindent
where $z_s$ are the poles of $f(z)$ in $z$ in the complex plane; $
Res f(z_s)$ is the residue of $f(z)$ at $z_s$ and we have assumed
the function $f(z)$ falls off at least as fast as
$1/|z|^{1+\epsilon}$ for large $\mid z \mid$. It will be convenient
to use:

\[   \tanh {\beta z_s \over 2} = 1- 2 n_f(z_s) \]

\noindent
where
\[
   n_f(z) = {1 \over e^{\beta z} +1 } \]

\noindent
is the usual Fermi-Dirac distribution function.

Rotating equations (2.11) and (2.12) into Euclidean space as
described above, we get:

$$
{1 \over 2 \lambda} = - {\partial V_{eff}^{(1)} \over \partial m^2}
=  \int {d^2 k \over (2 \pi)^2} {[-k_4^2 - k_1^2 + \mu^2 + M^2 - m^2]
\over D}
$$

$$
{1 \over \kappa} = - {\partial V_{eff}^{(1)} \over \partial M^2}
=  \int {d^2 k \over (2 \pi)^2} {[-k_4^2 - k_1^2 - \mu^2 - M^2 + m^2]
\over D}
$$

\noindent
where now the integral on $k_4$ is defined by eq.(\ref{eq:euc}),

\bigskip\noindent
where $D = [-k_4^2 - k_1^2 - M^2 + m^2 - \mu^2]^2 - 4[- m^2 k_4^2 +
\mu k_1^2 - m^2 k_1^2]$. There is no longer any need for an $i \epsilon
$
in the definition of $k_4$. Performing the sums over the Matsubara
frequencies we obtain the unrenormalized form of the equations
which are given by the same expression as the zero temperature ones
found earlier, with the replacements:

\begin{eqnarray}
  {1 \over k_{+}} \rightarrow && {1 \over k_{+}} ( 1- 2 n_f (k_{+}))
\nonumber \\
  {1 \over k_{-}} \rightarrow && {1 \over k_{-}} ( 1- 2 n_f (k_{-}))
\end{eqnarray}

As before we can integrate this to get the determinant correction to the
effective potential which in unrenormalized form is:

\begin{equation}
V^{(1)} (m,M) = -{1 \over 2 \pi} \int_0^{\Lambda} dk_1 [ k_+ + k_- + {2
\over \beta} \ln (1 + e^{-\beta k_+}) + {2 \over \beta} \ln (1 +
e^{-\beta k_-}) ]    \label{eq:v1mM2}
\end{equation}

\section{The case $\mu = T = 0$}

\bigskip
Renormalization of the effective potential is best discussed in the
context of the zero temperature and density sector of the theory
where we can define the renormalized coupling constant in terms of
the physical scattering of Fermions at a particular momentum scale.
This vacuum sector is interesting in its own right and we shall be
able, by analytic means, to derive the result that depending on a
parameter $\delta$ related to the
relative strengths of the two couplings the theory will be in one or
another broken phase and only in a mixed phase when $\delta=0$. Setting
$\mu =
T= 0$ we obtain

\begin{equation}
{\partial V_{eff}^{(1)} \over \partial m^2} = - {1 \over 4 \pi}
\int_0^{\Lambda}dk_1 [(1 + {M \over m}) {1 \over \sqrt{k_1^2
 + (m + M)^2}} + (1 - {M \over m}) {1 \over \sqrt{k_1^2 + (m
- M})^2]}
\end{equation}

\begin{equation}
{\partial V_{eff}^{(1)} \over \partial M^2} = - {1 \over 4 \pi}
\int_0^{\Lambda}dk_1 [(1 + {m \over M}) {1 \over \sqrt{k_1^2
 + (m + M)^2}} + (1 - {m \over M}) {1 \over \sqrt{k_1^2 + (m
- M})^2]}
\end{equation}

\bigskip\noindent
which is solved by

\begin{equation}
V^{(1)}(m, M) = - {1 \over 2\pi} \int_0^{\Lambda}dk_1 [\sqrt{k_1^2
 + (M + m)^2} + \sqrt{k_1^2
 + (M - m)^2} - 2 k_1] ~.
\end{equation}

\bigskip\noindent
This can be integrated to give the unrenormalized effective
potential:

\begin{eqnarray}
V_{eff}(m, M) &=& M^2 [{1 \over \kappa} - {1 \over 4\pi}] + m^2 [{1
\over 2\lambda} - {1 \over 4\pi}] \nonumber \\
&-& {1 \over 4 \pi} [(M + m)^2 \ln ({2 \Lambda \over M + m}) + (M -
m)^2 \ln ({2 \Lambda \over \mid M - m
\mid})] ~.
\end{eqnarray}

\bigskip\noindent
We renormalize by demanding that the renormalized couplings
$\kappa_R$ and $\lambda_R$ satisfy

\begin{equation}
{\partial^2 V_{eff} \over \partial B \partial
B^{\dagger}}\mid_{\stackrel{M
= M_0}{m = m_0}} = {4 \over \kappa_R}    \label{eq:vbb}
\end{equation}

\noindent
and

\begin{equation}
{\partial^2 V_{eff} \over \partial m^2}\mid_{\stackrel{M
= M_0}{m = m_0}} = {1 \over \lambda_R} ~.   \label{eq:vmm}
\end{equation}

\bigskip\noindent
Here $M = M_0$, $m = m_0$ designates an arbitrary renormalization
point on which the couplings will depend. Using these conditions to
solve for $\lambda$ and $\kappa$ in terms of $\lambda_R$ and
$\kappa_R$ yields the renormalized form of the effective potential:

\begin{eqnarray}
V_{eff} = m^2 [a &+& {1 \over 4 \pi} \ln \mid {M^2 - m^2 \over
\gamma_0}
\mid ] + M^2 [b + {1 \over 4 \pi} \ln \mid {M^2 - m^2 \over \gamma_0}
\mid ]
\nonumber \\
&+& {1 \over 2 \pi} m M \ln \mid {M + m \over M
- m} \mid   \label{eq:Veff1}
\end{eqnarray}

\bigskip\noindent
where $a$ and $b$ are the following constants:

\begin{eqnarray}
a &=& {1 \over 2 \lambda_R} - {3 \over 4 \pi} \nonumber \\ &~~&
\nonumber
\\ b &=& {1
\over
\kappa_R} - {1 \over 2 \pi} + {1 \over 8 \pi} {m_0 \over M_0} \ln
\mid {M_0 - m_0 \over M_0 + m_0} \mid     \label{eq:ab}
\end{eqnarray}

\noindent
and $\gamma_0 = \mid M_0^2 - m_0^2 \mid$.

Note that the renormalization we have just performed at $\mu = T =
0$ is also sufficient to remove all divergences from the effective
potential in the more general case of non-vanishing chemical
potential and temperature. The addition of $\mu$ and $T$ will only
result in finite corrections to the gap equations and therefore to
the vacuum values of $m$ and $M$. We shall return to this point in
section V.

For future reference we also want to consider the special
renormalization point relevant
for the sector  where there is chiral symmetry breaking
but no Cooper-pair gap when $\mu = T=0$. That is we will choose our 
renormalization point to be  the minimum of 
the potential which occurs in that case at 
\begin{equation}
m= m_F;  ~~~   M=0.
\end{equation}

For the choice $m_0= m_F,~~M_0=0$ , $a$ remains the same but $b$ reduces to
\begin{equation}
b= {1
\over
\kappa_R} - {3 \over 4 \pi }
\end{equation}
The renormalized coupling $\lambda_R$ takes on the particular value
$\pi$
and $V_{eff}$ simplifies to 

\begin{equation}
V_{eff} = M^2 ({1 \over \kappa_R} - {1 \over 4 \pi }) + { (m^2 + M^2 )
\over 4 \pi}~~
 (\ln \mid {M^2 - m^2 \over m_F^2} \mid -1)
+  {1 \over 2 \pi} m M~ \ln \mid {M + m \over M
- m} \mid
\end{equation}.

Here we want to point out that the quantity 
\begin{equation}
\delta \equiv b-a = {1 \over \kappa_R}-{1 \over 2 \lambda_R}={1 \over \kappa}-{1 \over 2 \lambda}
\end{equation}
so that $\delta$ is the same number before and after renormalization.

The gap equations are properly derived by differentiating $V_{eff}$
with respect to $B$ and $m$ and then setting these derivatives to
zero. Because $V_{eff}$ depends only on $B^{\dagger}B$ and $m^2$,
it will always be possible to have solutions with one of $m$ or $B$
or perhaps both set to zero. Differentiating eq.(\ref{eq:Veff1}) we obtain
the gap equations:

\begin{equation}
m [2a + {1 \over 2 \pi} + {1 \over 2 \pi} ~\ln {\mid M^2 - m^2 \mid
\over \gamma_0}] - {M \over 2 \pi} ~\ln \mid {M - m \over
M + m} \mid = 0  \label{eq:gap1}
\end{equation}

\noindent
and

\begin{equation}
M [b + {1 \over 4 \pi} + {1 \over 4 \pi} ~\ln {\mid M^2 - m^2 \mid
\over \gamma_0}] - {m \over 4 \pi} ln \mid {M - m \over M +
m} \mid = 0 ~.  \label{eq:gap2}
\end{equation}

\bigskip\noindent
The solutions $m = m^*$ and $M = M^*$ will give us the local
extrema of $V_{eff}$. The first of these equations is an identity
if $m
= 0$, and the second if $M = 0$. Also the values $m^*$ and $M^*$
that solve these equations are physical parameters that must be
independent of the renormalization scale $\gamma_0$. Thus these
equations tell us how $a$ and $b$ individually run with $\gamma_0$.
We note, however, that if we solve for the combination

\begin{equation}
\delta = b - a = {1 \over 4 \pi} [{m^{*2} - M^{*2}  \over  m^* M^*}]
~\ln
\mid {M^* - m^* \over M^* + m^*} \mid
\end{equation}

\bigskip\noindent
the scale $\gamma_0$ drops out. Therefore $\delta$ is a true
physical parameter in the theory; we shall see in the next section
that its value controls which of the two condensates $m$ and $M$
can exist. In the particular case where the minimum of the 
potential occurs when $m^* = m_F$ and $M^* = 0$ we have the simple
result:
\begin{equation}
\delta =  {1 \over \kappa_R} -{1 \over 2 \lambda_R }
= {1 \over \kappa_R}-{1 \over 2 \pi} .
\end{equation}

\bigskip
\section{Analysis of the gap equations}

It will be useful in the following to note that, at a solution of
the gap equations (3.12) and (3.13), the effective potential takes
the simple form

\begin{equation}
V_{eff} (m, M) = - {1 \over 4 \pi} (m^2 + M^2) ~. \label{eq:double}
\end{equation}

\bigskip\noindent
Our goal is to analyze all the solutions of the gap equations and
to find the one that produces the global minimum of $V_{eff}$. This
will then represent the true vacuum of the theory.

There are four types of solution to (\ref{eq:gap1}) and (\ref{eq:gap2}). The first is
simply to set $m = M = 0$, leading of course to $V = 0$. Clearly,
from (\ref{eq:double}) we see that if any other solution exists, $V = 0$ cannot
be the minimum of $V$. The second and third types are obtained by
setting $M = 0$, $m \neq 0$ and $m = 0$, $M \neq 0$ respectively.
If $M = 0$, then from (\ref{eq:gap1}), we have

\begin{equation}
m^2 = \gamma_0 ~ e^{ - (1 + 4 \pi a)}
\end{equation}

\noindent
so

\begin{equation}
V_0 (m, M = 0)= - {\gamma_0 \over 4 \pi} e^{-(1 + 4 \pi a)} \label{eq:vee2}
\end{equation}

\bigskip\noindent
(we shall use $V_0$ to denote values of $V_{eff}$ at solutions of
the gap eqn.). Likewise, if $m = 0, M \neq 0$, then from (\ref{eq:gap2})

\begin{equation}
M^2 = \gamma_0 ~ e^  { - (1 + 4 \pi b)}
\end{equation}

\begin{equation}
V_0 (m = 0, M) = - {\gamma_0 \over 4 \pi} ~e^{- (1 + 4 \pi b)} ~.
\label{eq:vee3}
\end{equation}

\bigskip\noindent
Thus we see that

\begin{equation}
V_0 (m = 0, M) < V_0 (m, M = 0) ~~~{\rm if} ~\delta < 0
\end{equation}

\noindent
and

\begin{equation}
V_0 (m, M = 0) < V_0 (m = 0, M) ~~~{\rm if} ~\delta  > 0 ~.
\end{equation}

The fourth case is when both $m$ and $M$ are non-vanishing. 
The analysis of this case is presented in the Appendix where it is shown that the solution with non-vanishing $m$ and $M$ always has $V_{eff}$ intermediate between the 
values of $V_{eff}$ associated with the two cases where one or the other of the condensates vanish.

We conclude that the global minimum of $V_{eff}$ has $M = 0, m \neq
0$ if $\delta > 0$, and $m = 0, M \neq 0$ if $\delta < 0$.

As we shall find later, the special point $\delta=0$ is the limit point
of the
line in $\mu,T$ space where there is a first order phase transition from
the phase with chiral symmetry breaking to the phase where there is only
superconductivity.

\section{Renormalized Effective Potential}
From eqn. (\ref{eq:v1mM2}) we can see that the corrections due to
non-vanishing temperature and density do not affect the ultraviolet
behavior of the integrand in the $k_1$ integral defining $V^{(1)}$.
Therefore, the renormalization that we have performed at $\mu = T =
0$ in section III suffices to remove the ultraviolet divergences
from the effective potential, and will allow us to send the cutoff
to infinity. It is perhaps worth recording the complete result
explicitly. We find, from eqns. (\ref{eq:vbb}) and (\ref{eq:vmm}), that

\begin{equation}
{1 \over 2 \lambda} = a + {1 \over 4 \pi} + X
\end{equation}

\begin{equation}
{1 \over \kappa} = b + {1 \over 4 \pi} + X
\end{equation}

\bigskip\noindent
where $a$ and $b$ are defined by eqn. (\ref{eq:ab}), and $X$ is a divergent
integral given by

\begin{eqnarray}
X &=& {1 \over 4 \pi} \int_0^{\Lambda} dk_1 [{1 \over \sqrt{k_1^2 +
(m_0 + M_0)^2}} + {1 \over \sqrt{k_1^2 + (m_0 - M_0)^2}}]
\nonumber \\
&=& {1 \over 2 \pi} ~[\ln ~({2 \Lambda \over \sqrt{\gamma_0}})]+
~{\rm terms ~which ~vanish ~as}~ \Lambda \rightarrow \infty ~.
\end{eqnarray}

\noindent
Thus the full renormalized effective potential may be written

\begin{eqnarray}
V_{eff} &=& \alpha_1 m^2 + \alpha_2 M^2 - {1 \over 2 \pi}
\int_0^{\infty} dk_1 [k_+ + k_- + {2 \over \beta} \ln~ (1 + e^{-\beta
k_+})
\nonumber \\
&+& {2 \over \beta} \ln~ (1 + e^{- \beta k_-}) \nonumber \\ &-&
2k_1
-({m^2 + M^2 \over 2}) ({1 \over \sqrt{k_1^2 + (m_0 + M_0)^2}} + {1
\over \sqrt{k_1^2 + (m_0 - M_0)^2}})] ~.
\end{eqnarray}

\bigskip\noindent
where $\alpha_1 = {1 \over 4 \pi} (1 + 4 \pi a)$ and $\alpha_2 = {1
\over 4 \pi} (1 + 4 \pi b)$. If $\alpha_1 < \alpha_2$, then at $\mu = T
=
0$ the vacuum has
$m^2 = m_F^2 \equiv \gamma_0 e^{-4 \pi \alpha_1} ~~{\rm and} ~~M^2
= 0.$ Here $m_F$ is the dynamically generated Fermion mass. It is
convenient to choose the renormalization scale so that $m_F^2 =
\gamma_0$. This entails setting  $\alpha_1 = 0$. 
Furthermore,
 we are free to choose $M_0 = 0$, so that $m_0 = m_F$. Then $V_{eff}$
takes
the form

\begin{eqnarray}
V_{eff} = \alpha_2  M^2 &-& {1 \over 2 \pi} \int_0^{\infty} dk_1
[k_+ + k_- + {2 \over \beta} (\ln ~(1 + e^{- \beta k_+}) + \ln ~(1
+ e^{- \beta k_-})) - 2 k_1 \nonumber \\ &-& (m^2 + M^2) {1 \over
\sqrt{k_1^2 + m_F^2}}] ~.
\end{eqnarray}

Here \[ \alpha_2 = \delta = {1 \over \kappa_R} - {1 \over 2 \pi} >0; ~~
\lambda_R= \pi,\]
as described above in Sec. III.

It is this branch of the theory that we are interested in as a model
for QCD, since QCD at zero temperature has a chiral condensate, but
does not have a Cooper-pair gap.  
We observe that if we set $M = 0$ in this expression, we obtain,
with $E = \sqrt{k^2 + m^2}$,

\begin{eqnarray}
V_{eff} (m^2, T, \mu) &=& {m^2 \over 4 \pi} [\ln~ {m^2 \over m_F^2}
- 1] \nonumber \\ &~& \nonumber \\ &-& {2 \over \beta}
\int_0^{\infty} {dk
\over 2
\pi} ~[\ln~ (1 + e
^{- \beta (E + \mu)})+ \ln (1 + e^{- \beta (E - \mu)})]
\end{eqnarray}

\bigskip\noindent
which is the effective potential for the Gross-Neveu model in
agreement with refs  \cite{ref:GN2} \cite {ref:inagaki}.  We will use
the analytic information
already known about the G-N model as a benchmark for our numerical
work below.

\bigskip In the opposite case $\alpha_2 <
\alpha_1$, we have, in the $\mu = T = 0$ vacuum, $m^2 = 0$ and $M^2
= \Delta^2
\equiv
\gamma_0 e^{- 4 \pi \alpha_2}$, where $\Delta$ is the dynamically
generated
gap. So we
choose  $\alpha_2 = 0, \alpha_1 > 0$, and $m_0 = 0$, $\Delta^2 =
\gamma_0 = M_0^2$. The effective potential becomes

\begin{eqnarray}
V_{eff} = \alpha_1 m^2 &-& {1 \over 2 \pi} \int_0^{\infty} dk_1
[k_+ + k_- + {2 \over \beta} (\ln ~(1 + e^{- \beta k_+}) + \ln ~(1
+ e^{- \beta k_-})) - 2 k_1 \nonumber \\ &-& (m^2 + M^2) {1 \over
\sqrt{k_1^2 +
\Delta^2}}] ~.
\end{eqnarray}

For this case, by choosing $\alpha_2 = 0$ we obtain:
\[ \kappa_R = 4 \pi, ~~-\delta= \alpha_1={1 \over 2 \lambda_R} - {1\over
2 \pi}
\ge 0.\]

\bigskip
When $m^2 = 0$, this expression gives us the effective potential at
finite temperature for the pure Cooper-pairing model considered in
\cite{ref:paper1}. Explicitly we have

\begin{eqnarray}
V_{eff} = {M^2 \over 4 \pi} [\ln {M^2 \over \Delta^2} - 1] - {2
\over \beta} \int_0^{\infty} {dk \over \pi} \ln~ [1 + e^{- \beta
\sqrt{k^2
+ M^2}}] ~.
\end{eqnarray}

\bigskip\noindent
Note that it is independent of the chemical potential, as was the
case at $T = 0$. 

\section{Phase structure of the Class of models}

\subsection{Cooper Pair Model}

The pure Cooper pair model  \cite{ref:paper1} has the property that the
chemical potential
is irrelevant and can be transformed away.  The form of the effective
potential is exactly the same as that for the Gross Neveu model at zero
chemical potential  with $M$
replacing $m$ and the gap $\Delta$ replacing $m_F$.   Thus in leading
order large-N there is a second order
phase transition to the unbroken mode at a critical temperature which
can be determined by the high temperature expansion.  
For $T >> M$ we can expand the integral in eq.
(5.8) to obtain \cite{ref:GN2}

\begin{eqnarray}
V_{eff} = {M^2 \over 2 \pi} [\ln~ ({\pi T \over \Delta}) - \gamma]
~,
\end{eqnarray}

\bigskip\noindent
where $\gamma$ is Euler's constant. The minimum of this function
occurs at $M = 0$, which means that the condensate vanishes for
large $T$, as expected. The critical temperature is that temperature for
which
\begin{equation}
 \ln~ ({\pi T \over \Delta}) - \gamma =0  \rightarrow T_c= {\Delta \over
\pi} e^{\gamma}.
\end{equation}

The same critical temperature was obtained in another variant of the
Gross-Neveu
model which had a superconducting phase \cite{ref:GNsup}, so that this
temperature seems ubiquitous
in 4-Fermi models in 1+1 dimensions.

\subsection{Gross-Neveu sector}
As is well known, the Gross Neveu Model has
spontaneous symmetry breaking at zero chemical potential and
temperature.
At zero temperature, the symmetry is restored at finite chemical
potential at
a critical value of $\mu$ \cite{ref:GN2} \cite{ref:minakata}. This transition is first order.  At zero
chemical
potential the system undergoes a second order phase transition  to the
unbroken symmetry phase as we increase the temperature.   Thus at some
point in the phase diagram there is a tricritical point.  For this model
we have performed both high and low temperature expansions of the
leading order in $1/N$  potential which is given by:
\begin{eqnarray}
V_{eff} (m^2, T, \mu) &=& {m^2 \over 4 \pi} [\ln~ {m^2 \over m_F^2}
- 1] \nonumber \\ &~& \nonumber \\ &-& {2 \over \beta}
\int_0^{\infty} {dk
\over 2
\pi} ~[\ln~ (1 + e
^{- \beta (E + \mu)})+ \ln (1 + e^{- \beta (E - \mu)})]
\end{eqnarray}

 In the high temperature regime, using methods similar to those
used for Bose condensation \cite{ref:Bose} we obtain:
\begin{equation}
V_{eff} (m^2, T, \mu)= {m^2 \over 4 \pi} ( \ln ~ {
T^2 \over T_c^2} + {7 \over 2} {\zeta(3) \over \pi^2  T^2}(\mu^2 + {m^2
\over
4}) ) \end{equation}
which leads to the relationship: 
\begin{equation} 
T_c= {m_F \over \pi} \exp [\gamma - {7 \mu^2 \zeta(3) \over 4 \pi^2
T_c^2}]
\end{equation}
which at $\mu$= 0 gives the same critical temperature as for the 
Cooper pair model, however with $m_F$ replacing $\Delta$.
At small $\mu^2$ one has approximately
\begin{equation} 
T_c= {m_F \over \pi} e^{\gamma} [1- {7 \mu^2 \zeta(3) \over 4 \gamma
m_F^2  e^{\gamma}}]
\end{equation}

In the low temperature regime we want an analytic
expression
for the effective potential which would enable us to determine values
of $\mu$
and $T$ at which the first order transition  occurs.    At zero $T$ the
modification to the effective potential due to the chemical potential is
only
in the  region $m \leq \mu$.  The standard low temperature expansion
used in
Bose Condensation \cite{ref:Bose} unfortunately only gives the finite
temperature corrections  when $ \mu \leq m$ and thus is not very
relevant to
the question we want to  answer.  
To obtain an approximate analytic expression valid in the opposite
regime
$ m \leq \mu$ pertinent to the first order phase transition, we resort
to a crude approximation which captures the relevant physics.
That is we make an approximation to the Fermi-Dirac distribution  
function that allows us to perform all the integrals. First we
rewrite the  derivative of the potential in the form:
\begin{equation}
{\partial V \over \partial m} = {m \over 2 \pi} \ln {m^2 \over m_F^2} +
{m \over \pi} \int_0^{\infty} {dk \over E} [2 - \tanh{E+ \mu \over 2T} -
\tanh{E- \mu \over 2T} ],
\end{equation}
where $ E = \sqrt{k^2+m^2}$. 
and then replace the function  $\tanh({E-\mu \over T})$  using the
straight
line interpolation: 
\begin{equation}
 \tanh(x) \rightarrow \{ 1 ~{\rm if}~  x>2;~ -1 ~{\rm if}~x < 2;~   ~ x ~{\rm if}~ |x| \leq 2 \}. \label{eq:stline}
\end{equation}
This has the correct behavior as $T \rightarrow 0$ and captures the physics of
the
broadening of the Fermi surface.
\
At $T=0$, the effect of the chemical potential is the most dramatic.
Since in that limit   $\tanh x = \epsilon(x)$, we get immediately 
that 
\begin{eqnarray}
{\partial V \over \partial m}&& = {m \over 2 \pi} \ln {m^2 \over m_F^2}
+ 
{m \over \pi} \int_0^{\sqrt{\mu^2-m^2}} {dk \over E} \Theta(\mu -m )
\nonumber
\\
 && = {m \over 2 \pi} \ln {m^2 \over m_F^2} + 
{m \over \pi} \Theta(\mu -m)  \{ \ln(1+ \sqrt{1-m^2/\mu^2}) -{1 \over 2}
\ln{m^2 \over \mu^2} \}
 , \end{eqnarray}
This can be integrated to give the result that for $m \leq \mu$
the effective potential is given by:
\begin{equation}
V_{eff} = {1 \over 4 \pi}  \{  m^2 ( 2 ~ \ln [{\mu + \sqrt{\mu^2 - m^2}
\over m_F}] - 1) - 2 \mu \sqrt{\mu^2 - m^2} \} + C(\mu) 
\end{equation}
whereas, for  $m > \mu$ the effective potential is equal to its $\mu = 0$
value, namely
\begin{equation}
V_{eff} ={m^2 \over 4 \pi} [\ln~ {m^2 \over m_F^2}
- 1] + C(\mu)
\end{equation}
The arbitrary integration constant can be eliminated by choosing
$V_{eff}(m=0)
=0,$which yields
\begin{equation}
 C(\mu) = {\mu^2 \over 2 \pi} \label{eq:constant}
\end{equation}

  For all $T$ we can use the approximation in Eq. (\ref{eq:stline})  to
perform all the integrals explicitly. Doing this,  we obtain an
approximation to
the exact phase structure in the regime where there is a first order 
phase transition as is shown in Fig. 1. In that figure we also include
the high temperature analytic result.  Our
analytic  calculation gives us an approximate value for the tricritical point which
separates the regime between the first and second order phase
transitions: $
{\mu_c \over m_F} = .661 ~~~  {T_c  \over m_F} = .31 $
compared to the ``exact'' numerical result as for example found in Ref. 
\cite{ref:GN2}
\begin{equation}
 {\mu_c \over m_F} = .608  , ~~~
~~~  {T_c  \over m_F} = .318 . \label{eq:tri}
\end{equation}

\subsection{Full Phase Structure}
The phase structure is quite different depending on whether we choose
the 
case $\delta > 0  $ which has chiral symmetry breaking in the vacuum,
or  $\delta < 0 $  where there is Cooper pair formation in the
vacuum.
In the regime where $\delta > 0 $, the phases of this model
are quite similar to QCD as shown in  figures 2  with the value
of $\alpha_2=\delta = {1 \over 2 \pi}$.  In the
vacuum there is chiral symmetry breakdown. As we increase the chemical
potential at low temperatures there is a first order phase transition
into
a phase with Cooper pairs. At and near the phase  transition line there
can be coexistence of the two separate phases, one with Cooper pairs
and one with a chiral condensate which breaks chiral symmetry.  
For the range 
\begin{equation}
 \delta> {1.13097 \over 4 \pi} = \delta_c
\end{equation}
the theory will have a tricritical point at the value given by Eq. (\ref{eq:tri}), so that the regime where there
is chiral symmetry breakdown will, for chemical
potentials below the tricritical value undergo a second order phase transition
at large temperatures.
For values of the chemical potential between the tricritical value and
the value for the first order transition to the superconducting phase
(determined below), the phase
transition from the chirally broken mode to  the unbroken mode will be first  order at large temperatures.

As we move to higher values of the chemical potential, for sufficiently low temperatures the system exists in a superconducting phase with nonzero
gap.  As we increase the temperature at fixed large chemical potential, the
system undergoes a second order transition into the unbroken mode, with
the critical temperature depending only on $\delta$ and not $\mu$. This
dependence is displayed in figure 3. 
$T_c$ reaches
the tricritical
 value $T_{c}/m_F = .318$ when $\delta =\delta_c $. 
Figure 2 is in the regime where $\delta > \delta_c$ so that it displays
a tricritical point.  
 For values
of $\delta < \delta_c$, the chirally broken phase only can be restored via
a second order phase transition. This case is illustrated in fig.4  which
is for $ \delta = {1 \over 4 \pi}< \delta_c$.  In between the chirally
broken and superconducting phases is
a coexistence curve.  The intersection of this curve with the line  $T=0$ can be
determined
as a function of $\delta$ which we will shall do below.  The existence
of two phases having the same energy is shown in the 
3D plot of
the effective potential as a function of $m,M$ in fig.5 and in the
two dimensional slices of this figure shown in figs. 6 and 7. 
The particular case displayed  is for
 $\alpha_2 =\delta = {1 \over 4 \pi} $.  All the plots are for 
$T= 0.02, \mu = 0.56$
which is numerically determined  to lie along the first order line
separating the chiral condensation phase from the Cooper condensation 
phase.
 As $\delta$
approaches zero, the coexistence curve approaches $\mu=0$ and after that
one no longer has a phase with chiral symmetry breakdown.

Infinitesimally to either side of the coexistence curve we are at two separate minima of the potential. 
In each phase the minimum takes place with the value of the other condensate
mass equal to zero. Thus the condition defining the coexistence curve is
\begin{equation}
V[m=m_F, M=0] = V[m=0, M= M^*] 
\end{equation}   
The value of $M^*$ is chosen to minimize $V[0,M]$ for a given value
of $\mu, \delta$.
Recall that at $T=0$ the effective potential is given by:
\begin{eqnarray}
V_{eff}&=&\delta~ M^2 -\frac{1}{2\pi} \int_0 ^{\infty} dk
[
 {\sqrt{{k^2} + {m^2} + {M^2} + {\mu^2} + 
      2\,{\sqrt{{m^2}\,{M^2} + \left( {k^2} + {m^2} \right)
\,{\mu^2}}}}}
\nonumber\\
& &+
 {\sqrt{{k^2} + {m^2} + {M^2} + {\mu^2} - 
      2\,{\sqrt{{m^2}\,{M^2} + \left( {k^2} + {m^2} \right)
\,{\mu^2}}}}}
-2k
- {\frac{{m^2} + {M^2}}{{\sqrt{m_F^2 + {k^2}}}}}] +C(\mu)
\end{eqnarray}

We notice that when $M=0$, this potential becomes that of the
Gross-Neveu
model.
Thus if we choose $V_{eff} (m=0,M=0) = 0$, then again $C(\mu)= {\mu^2
\over
2 \pi}. $, as in eq. (\ref{eq:constant}).

At $T=0$ it is possible to analytically determine the value of the
chemical
potential  as a function of $\delta $ as well the value of $M$ at the
minimum. 
On the left hand side of the coexistence we need to evaluate the GN
effective
potential in the regime where $m=m_F   > \mu$, since the phase
transition to the
superconducting phase always occurs in that regime. Thus we have 
\begin{equation} 
V_{eff}(m=m_F,0) = -\frac{m_F^2}{4\pi}+\frac{\mu^2}{2\pi}
\end{equation}
On the right hand side we need to evaluate the zero temperature
effective
potential for $m=0, M=M^*$.
We have on the Cooper condensation side,
\begin{eqnarray}
V_{eff}(m=0,M)=
\delta  M^2+\frac{M^2}{4\pi}(\ln {M^2 \over m_F^2}  -1)
\end{eqnarray}
The quantity $M^*$ is determined by that value of $M$ that minimizes
this function, namely
\begin{equation}
{\partial V_{eff}(m=0,M) \over \partial M} =0 \rightarrow \delta + {\ln
{M^2 \over m_F^2}
\over 4 \pi} =0.
\end{equation}
or
\begin{equation}
M^{*2}  = m_F^2 e^{-4 \pi \delta}
\end{equation}
Inserting this value into the equation equating the value of the
potential on 
both sides of the phase transition we then obtain the critical value of
the 
chemical potential
\begin{equation}
 {\mu^2 \over m_F^2} = { 1- e^{-4 \pi \delta}\over 2 }.
\end{equation}

\section{Conclusions}
In this paper we have analyzed a 1+1 dimensional model possessing $O(N)$
flavor symmetry and discrete chiral symmetry, and have found a phase structure
remarkably similar to that conjectured for 2-flavor QCD. We have derived the
general forms for the effective potential in leading order in ${1 \over N}$.
 We have analyzed the case $\mu = T = 0$ analytically,
showing how the phase structure is governed by the renormalization group
invariant $\delta$. For $\mu = T = 0$ this structure is remarkably symmetric
in the two condensates $m$ and $M$. We have performed careful numerical
analysis of the integrals involved in the determination of the effective
potential and have determined the  dependence of $V_{eff}$ 
on the parameters $\delta, \mu$ and $T$.  What we have found
is that when there is chiral symmetry
breakdown in the vacuum sector  ( $\delta  > 0$), there are at least three different
regions. In the low  temperature regime, as we increase the chemical
potential 
there is a first order phase transition to a regime which has a Cooper
pair gap (superconductivity) but no chiral symmetry breakdown. Along and
near the 
phase transition line, there is a regime where the two phases coexist
like
ice and water. At $T=0$ we explicitly determine the value of the chemical
potential at which this occurs and also the value of the Cooper pair
gap as a function of $\delta$.  At
high enough temperatures both symmetries are restored.  In particular
if $\delta > \delta_c = {1.13097 \over 4 \pi}$, then there is also a tricritical point so that depending on the value of $\mu$ the phase transition
out of the chirally broken phase will be either first or second order.   We illustrated the phase structure of this model by showing the phase
diagram of this model as a function of temperature and chemical potential for
representative values of $\delta$. We also plotted the effective potential at a
representative place where there is phase coexistence.
 In the opposite
case $\delta < 0$ one finds that in the vacuum sector the theory
has a Cooper pair gap but no chiral symmetry breaking. In that case the theory 
has a transition at high temperatures to the unbroken mode where
the gap goes to zero. 

Using this toy model we intend to study how the phase transition from the high
temperature to low temperature regime proceeds in time during an
expansion
of an initial Lorentz contracted  disc of quark matter starting  from various initial conditions related to different points on
this phase diagram. We hope to determine how various correlation
functions depend on the initial conditions of a scattering experiment, assuming
that it produces an initial state in local chemical and thermal equilibrium somewhere on the phase diagram we obtained in this paper.


\bigskip\noindent
{\bf Acknowledgements}

We wish to thank Gregg Gallatin for interesting conversations.
 The research of AC is
supported in part by DOE grant DE-FG02-92ER-40704. The research of 
FC and AS is supported by the DOE. In addition, AC
and HM are supported in part by the Grant-in-Aid for International
Scientific Research No. 09045036, Inter-University Cooperative
Research, Ministry of Education, Science, Sports and Culture of
Japan. This work has been performed as an activity supported by the
TMU-Yale Agreement on Exchange of Scholars and Collaborations. FC
and HM are grateful for the hospitality of the Center for
Theoretical Physics at Yale. HM, AC, and WM are grateful for the
hospitality of the theory group at Los Alamos. 

\newpage

\section*{Appendix A}


In this Appendix we give the details for determining that the
relative minimum which has both condensates is always between 
the two minima which have only one condensate.
When both $m$ and $M$ are non-vanishing, it is
then convenient to define $\rho = {M \over m}$ and to combine the
gap equations in the form

\begin{equation}
\delta = (1- \rho^2) [b + {1 \over 4 \pi} + {1 \over 4 \pi} \ln ({m^2
\mid {\rho^2 - 1}\mid \over \gamma_0})]  \label{eq:del1}
\end{equation}

\noindent
and

\begin{equation}
\delta = {1 \over 4 \pi} {(1 -\rho^2) \over \rho} ~\ln~ \mid {\rho - 1
\over \rho + 1} \mid ~.  \label{eq:del2}
\end{equation}

\bigskip\noindent
Both these equations are even in $\rho$, so we may take $\rho > 0$
for convenience. Eqn. (\ref{eq:del2}) tells us immediately that if $\delta <
0$, $0 < \rho < 1$, and if $\delta >  0$, $\rho > 1$. Furthermore,
the r.h.s. of (\ref{eq:del2}) is bounded between $- {1 \over 2 \pi}$ and ${1
\over 2 \pi}$. Hence we conclude: ~If $\mid \delta \mid > {1 \over 2
\pi}$
 there is no solution with both $m$ and $M$ non-vanishing. If $\mid
\delta
\mid < {1 \over 2 \pi}$,
 there is such a solution, with the property that $m > M$ if $\delta >
0$
and $M < m$ if $\delta < 0$.

It remains to decide whether $V_0(m, M)$ can be the global minimum.
To this end, it is convenient to re-express the gap equations once
more in the following form:

\begin{equation}
- (1 + 4 \pi a) = \ln ~{m^2 \mid \rho^2 - 1 \mid \over \gamma_0} - \ln
\{
~\mid {\rho - 1 \over \rho + 1} \mid^{\rho} \}
\end{equation}

\noindent
and

\begin{equation}
- (1 + 4 \pi b) = \ln ~{m^2 \mid \rho^2 - 1 \mid \over \gamma_0} - \ln
\{
~\mid {\rho - 1 \over \rho + 1} \mid^{{1 \over \rho}} \}~.
\end{equation}

\bigskip\noindent
From these, making use of eqs. (\ref{eq:double}), (\ref{eq:vee2}) and
(\ref{eq:vee3}), we immediately obtain

\begin{equation}
V_0 (m = 0, M) = - {\gamma_0 \over 4 \pi} ~e^{- (1 + 4 \pi b)} =
g_1(\rho) V_0(m, M)
\end{equation}

\noindent
and

\begin{equation}
V_0 (m, M = 0) = - {\gamma_0 \over 4 \pi} ~e^{- (1 + 4 \pi a)} =
g_2(\rho) V_0(m, M)
\end{equation}

\noindent
where

\begin{equation}
g_1(\rho) = {(1 + \rho)^{1 + {1 \over \rho}} \mid 1 - \rho
\mid^{1
- {1 \over \rho}} \over 1 + \rho^2}
\end{equation}

\noindent
and

\begin{equation}
g_2(\rho) = {(\rho + 1)^{\rho + 1} \mid \rho - 1
\mid^{1
-  \rho} \over 1 + \rho^2} ~.
\end{equation}

\bigskip\noindent
Eq. (7.5) is the relevant comparison if ${1 \over 2 \pi} < \delta
<  0$ and $0 < \rho < 1$, whereas eq. (7.6) is relevant for $0 <
\delta
<  {1 \over 2 \pi}$ and $\rho > 1$.

We observe, however, that $g_2({1 \over \rho}) = g_1(\rho)$, so
both cases reduce to the following: ~if we can show that $g_1(\rho)
> 1$ in the range $0 < \rho < 1$, then $V_0(m, M)$ is never the global
minimum
(recall that the $V_0's$ are all $< 0$). On the other hand, if
$g_1(\rho) < 1$ in this range, it will be possible to have $V_0(m,
M)$ be the global minimum.

To settle this question, write $g_1 = e^h$, with

\begin{eqnarray}
h(\rho) &=& (1 + {1 \over \rho}) ~\ln~ (1 + \rho) + (1 - {1 \over
\rho}) ~\ln~ (1 - \rho) - ~\ln~ (1 + \rho^2) \nonumber \\
&=& \ln~ [{1 + \rho \over 1 + \rho^2}] + {1 \over \rho} ~\ln~ (1 +
\rho) + (1 - {1 \over \rho}) ~\ln~ (1 - \rho) ~.
\end{eqnarray}

\bigskip\noindent
In the range of interest, $\rho^2 < \rho$, so the r.h.s. is a sum
of positive terms. Hence $h(\rho) > 0$ and $g_1(\rho) > 1$.

We conclude that the global minimum of $V_{eff}$ has $M = 0, m \neq
0$ if $\delta > 0$, and $m = 0, M \neq 0$ if $\delta < 0$.

\newpage
%
%

\begin{figure}
\epsfxsize = 3. in
   \centerline{\epsfbox{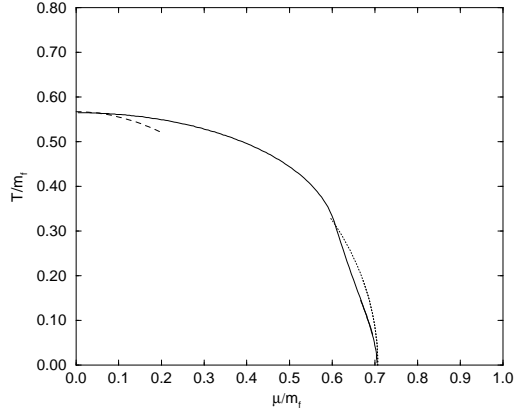}}
     \caption{Phase diagram for the Gross-Neveu model. Partial lines
are results of the High and  (approximate) Low Temperature expansions.
Continuous line is the numerical result.}
   \label{fig:Gross-Neveu}
\end{figure}


\begin{figure}
\epsfxsize = 3. in
   \centerline{\epsfbox{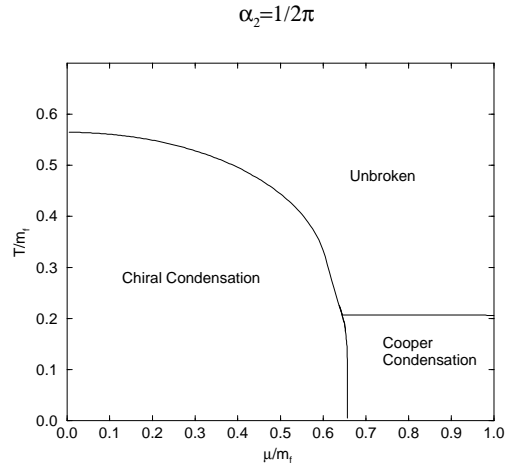}}
     \caption{Phase Structure at $\delta= {1 \over 2 \pi}$. Tricritical point
is at $T/m_F = .318, \mu/m_F = .608$.}
   \label{fig:alpha2_pi/2}
\end{figure}


\begin{figure}
\epsfxsize = 3. in
   \centerline{\epsfbox{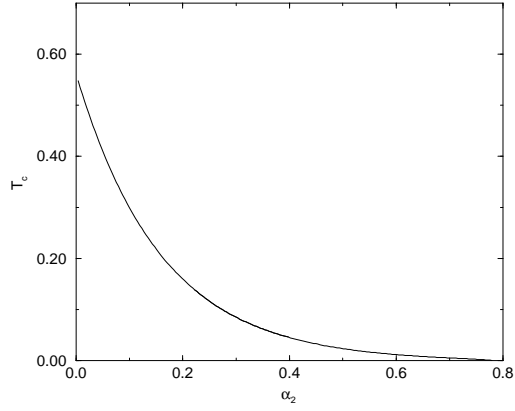}}
     \caption{Critical Temperature for the Superconducting Phase as a
function
of $\alpha_2=\delta$}
   \label{fig:TCvA}
\end{figure}

\begin{figure}
\epsfxsize = 3. in
   \centerline{\epsfbox{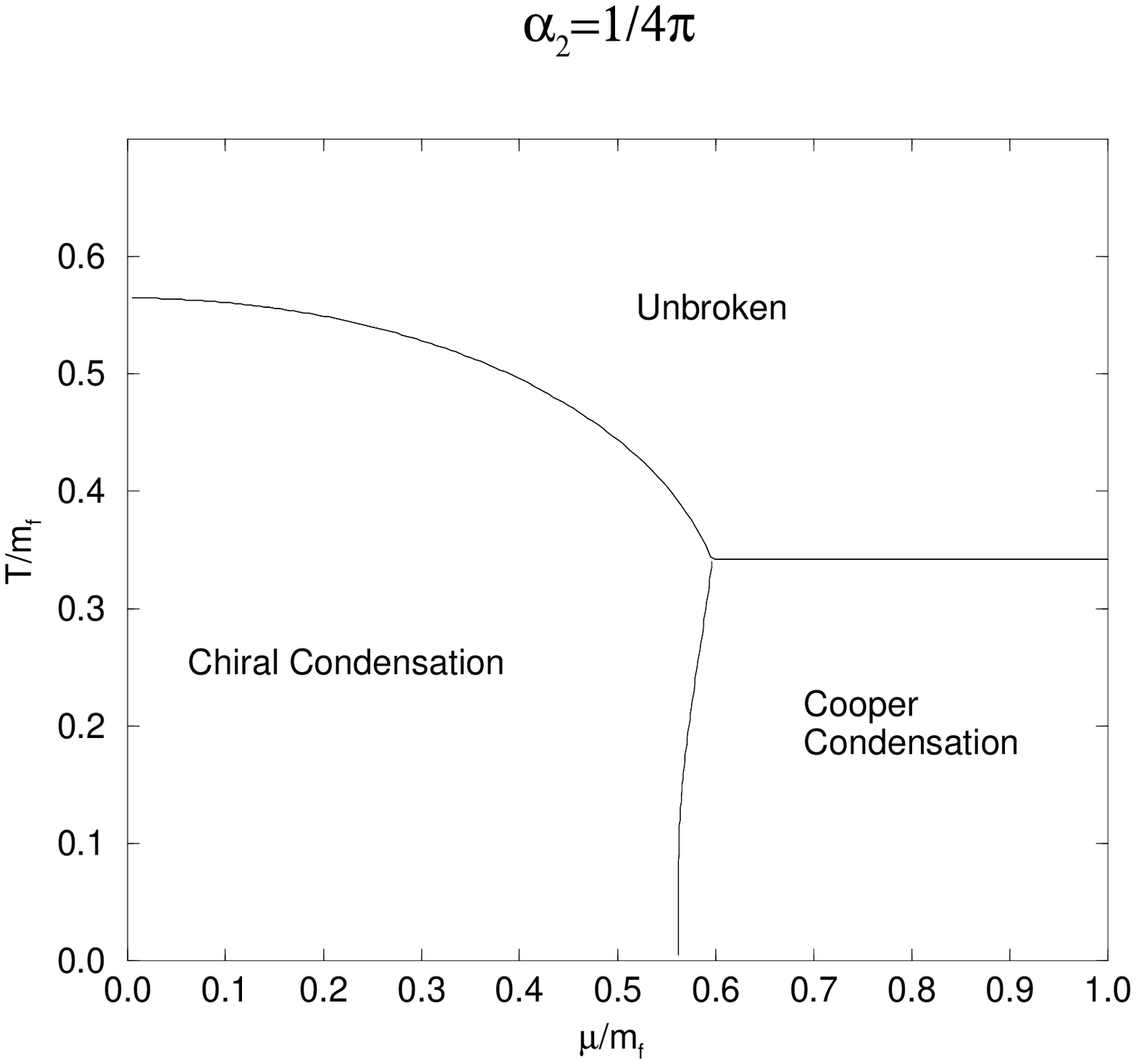}}
     \caption{Phase Structure at  $\delta= {1 \over 4 \pi}$}
   \label{fig:alpha2_pi/4}
\end{figure}

\begin{figure}
\epsfxsize = 3. in
   \centerline{\epsfbox{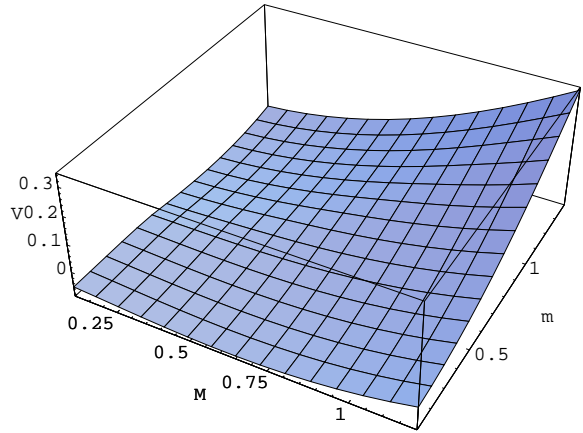}}
     \caption{Phase Coexistence Effective Potential at $T=0.02, \mu =
0.56
, \delta={1\over 4 \pi}$}
\end{figure}

\begin{figure}
\epsfxsize = 3. in
   \centerline{\epsfbox{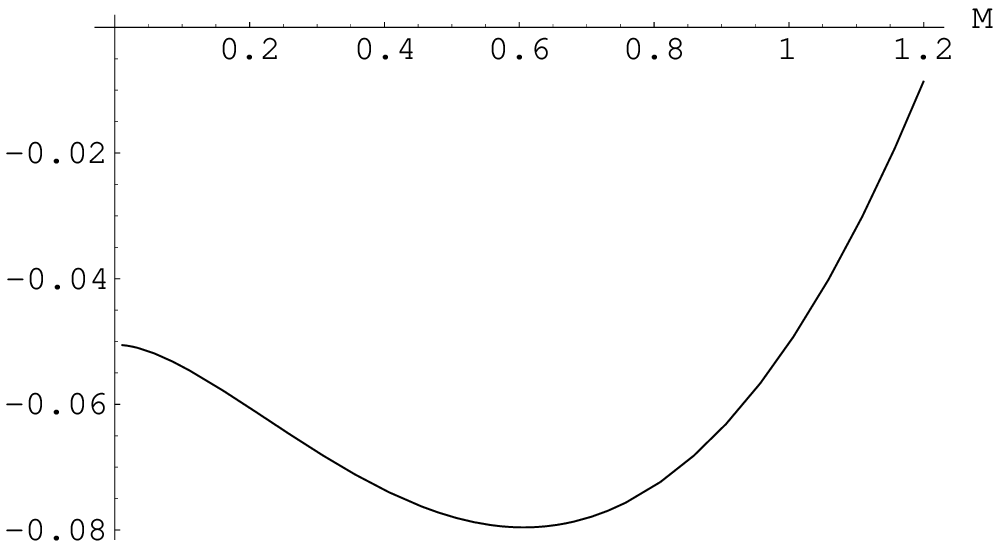}}
     \caption{Phase Coexistence Effective Potential as a function
of $M$ for $m=0$}
\end{figure}

\begin{figure}
\epsfxsize = 3. in
   \centerline{\epsfbox{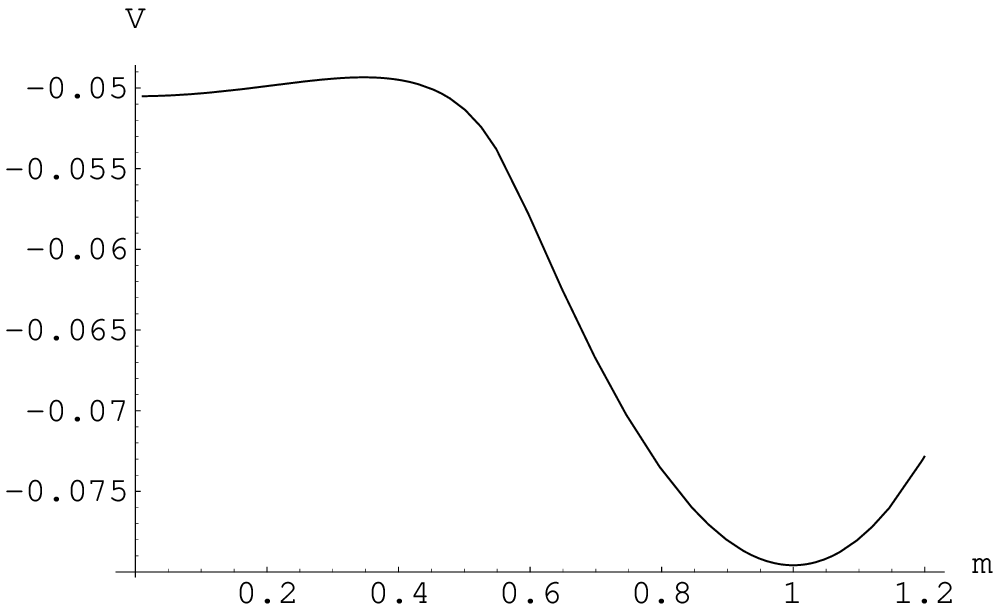}}
     \caption{Phase Coexistence Effective Potential as a function
of $m$ for $M=0$}
\end{figure}
\end{document}